\documentclass[10pt]{article}
\usepackage[varg]{txfonts}
\usepackage{epsf,rotating,latexsym,amssymb} 
\usepackage{epsfig}   
\usepackage{epstopdf} 
\usepackage{rotating,latexsym,amssymb}           
\usepackage{mathptmx} 
\usepackage{graphicx}
\usepackage{times}
\usepackage{natbib}
\begin{document}
\bibliographystyle{natbib}
%
%


\noindent
\Large{\bf FORMATION OF THE THEBE EXTENSION IN THE  RING SYSTEM OF JUPITER}

\bigskip

\noindent
N. Borisov$^{1}$ and H. Kr\"uger$^{2}$ (krueger@mps.mpg.de)
\medskip

\noindent
(1) {IZMIRAN, 142190, Kaluzshskoe Hwy 4,Troitsk, Moscow, Russia} \\
(2) {MPI for Solar System Research, 37077 G\"ottingen, Germany}



\bigskip
\bigskip

\section*{Abstract}
Jupiter's tenuous dust ring system is embedded in the planet's inner magnetosphere, and -- among other 
structures -- contains a very tenuous protrusion called the Thebe 
extension. In an attempt to explain the existence of this swath of particles beyond Thebe's orbit, 
\citet{hamilton2008} proposed that the dust particle motion is driven by a shadow resonance caused by
variable  dust charging on the day and night side of Jupiter. However, the model by \citet{divine1983} together 
with recent 
observations by the Juno spacecraft
indicates a warm and rather dense inner magnetosphere of Jupiter which implies that
the mechanism of the shadow resonance does not work. Instead we argue that dust grains ejected from
Thebe due to micrometeoroid bombardment become the source of dust in
the Thebe extension. We show that large (grain radii of a few micrometers up to 
multi-micrometers) charged dust grains having
significant initial velocities oscillate in the Thebe extension. 
Smaller charged grains (with sub-micrometer radii) ejected from Thebe do not spend
much time in the Thebe extension and migrate into the Thebe ring. At the same time, if such grains
are ejected from larger dust grains in the Thebe extension due to fragmentation, they continue to oscillate
within the Thebe extension for years. We argue that fragmentation of large
dust grains in the Thebe extensions
could be the main source of sub-micrometer grains detected in the Thebe extension.

{\it Keywords:} Thebe; Cosmic Dust; Space Plasma; Electric fields,  Jupiter ring system, Planetary Rings

\section{Introduction}

Jupiter has a highly structured dust ring system that extends along the planet's equatorial plane.
It consists of the main ring, the halo, two
gossamer rings and the Thebe extension \citep{burns1984a,showalter1987,burns2004}.
This dust system  was investigated
by the space missions Voyager 1,2, Galileo, Cassini and New Horizons 
\citep{smith1979a,owen1979,ockert-bell1999,porco2003,burns2004,showalter2007,throop2016}
 and also by
telescopes in space (Hubble) and from the Earth (Keck) 
\citep{depater1999,depater2008,showalter2008}.

Usually it is assumed that the gossamer rings are formed by dust grains ejected from
the surfaces of the corresponding moons (Thebe and Amalthea) due to micrometeorite bombardment 
\citep{krivov2002a,krivov2003,dikarev2006}. Recently it has been argued that a supplementary mechanism of dust
grain ejection from the polar regions of Thebe and Amalthea, i.e. electrostatic lofting, contributes
to the formation of the gossamer rings \citep{borisov2020}. At the
same time the mechanism of the  formation of the Thebe extension is not so clear.
Indeed, it is known that  Jupiter's corotational radial electric field acts on dust
grains in the planet's inner magnetosphere \citep{horanyi1991}. The corresponding
electric force for negatively charged grains in the gossamer rings
is directed inwards (towards Jupiter) for radial distances $r > R_{\rm synch}$, where
$R_{\rm synch}=2.27\,\mathrm{R_J}$ is the radius of the synchronous orbit, i.e. the orbit at which the
angular frequency of a body orbiting Jupiter  
is equal to the angular frequency of
Jupiter's rotation ({$\mathrm{R_J=71,492\, km}$ is the radius of Jupiter).
On the contrary, positively charged grains are pulled
outwards  by this electric field (into the Thebe extension).

The sign and the magnitude of the electric charge on a dust grain in  Jupiter's shadow is
determined only by the plasma parameters in the inner magnetosphere of Jupiter, 
and additionally on the sun-lit side by the action of the solar UV radiation. Unfortunately,
real plasma parameters of  Jupiter's inner magnetosphere are still not well-known.
Earlier in the theoretical investigations devoted to dust grains dynamics in the gossamer rings
the authors often estimated the electron
concentration in the inner magnetosphere of Jupiter as of the order of $N_e
\sim 1\,\mathrm{cm^{-3}}$ or even less, and the electron temperature of a few electron volts, see, e.g.  
\citet{horanyi1991,hamilton2008}.  In such a plasma environment, the
electric potential on dust grains on the sun-lit side can be positive,
thus driving such charged dust grains outwards. This mechanism was suggested by  
\citet{hamilton2008} to explain
the formation of the Thebe extension. At the same time, according to the
well-known theoretical model of thermal plasma in the inner magnetosphere of Jupiter by \citet{divine1983},
the electron concentration in the vicinity of Thebe is  predicted to be
$N_e\approx 50\,\mathrm{cm^{-3}}$ and the electron temperature $T_e\approx 45\,$eV. This model is 
based on Pioneer and Voyager data  and was confirmed
later by \citet{garrett2005} (only slightly corrected). For such large values of the electron concentration
and temperature the electric charges on dust grains are definitely negative
and rather high, both on the sun-lit side and in the planet's shadow.
Furthermore, for such large concentrations of thermal electrons
the variation of the electric potential
on the sun-lit side and in the shadow of Jupiter should be almost  negligible
(as an example how the electric potential changes with the growth of the plasma concentration
see \citet{horanyi1991}, their Fig.~5). As a result, assuming that the Divine model is correct, 
the above mentioned  explanation of the Thebe
extension does not work,  and a new physical
mechanism for its formation has to be developed.

In relation with the  model  by \citet{divine1983} two aspects should be emphasized.
First, new experimental data obtained by the Juno  mission at $R\approx 10\,\mathrm{R_J}$ clearly
demonstrate that the concentration of ions grows rather quickly towards smaller
distances from Jupiter \citep{kim2020}. Second, it is expected that later on (at the end
of its mission) Juno will provide us with the real concentrations of charged particles at
$R\approx 3\,\mathrm{R_J}$.

According to the experimental data,
the Thebe extension is a very faint structure (dust concentration is smaller than in the gossamer
rings, see \citet{burns1999,ockert-bell1999,krueger2009b}). This means that possibly not
all dust grains ejected from Thebe penetrate into the Thebe extension but
only some fraction. The radial range of the Thebe extension
as seen in images \citep{ockert-bell1999} comprises at least approximately $0.64\,\mathrm{R_J}$ (from the
orbit of Thebe  at $3.11\,\mathrm{R_J}$ out to $3.75\,\mathrm{R_J}$) which is in agreement with in situ dust measurements by the Galileo
spacecraft \citep{krueger2009b}.

In this paper we take into account that dust grains lofted from the  moons' surface can have
rather high initial velocities. Velocities of the ejected dust
grains and their directions of propagation were investigated by various authors, e.g,
\citet{gault1963,eichhorn1978b,okeefe1985,koschny2001a,sachse2015}, and a recent review by \citet{szalay2018}. 
Laboratory  measurements together with numerical simulations 
were applied to obtain the distributions
of masses, velocities and their directions as functions of a meteorite (micrometeorite) parameters. Various authors
obtained quite different maximum ejection speeds, ranging from approximately $\mathrm{700-800\,m\,s^{-1}}$ up to 
$\mathrm{20\,km\,s^{-1}}$. 

In addition to these so far inconclusive results based on laboratory measurements, 
another line of evidence comes from  in situ dust measurements in the Jovian system: 
Impact ejecta dust clouds were directly measured at the Galilean moons \citep{krueger1999d,krueger2003b}.
These tenuous clouds were detected in the vicinity of these  moons up to  an altitude of approximately eight 
satellite radii. It  implies that the particles must have been ejected from the surfaces of these moons
with speeds not too far below 
their escape speeds, which are in the range of approximately $\mathrm{2.5\,km\,s^{-1}}$. Furthermore, the Galileo
detector measured a  very tenuous dust ring around the planet in the region between the Galilean moons. These 
particles were interpreted (at least partially) as being ejecta from these moons \citep{krivov2002a}. Therefore, 
they must have reached ejection speeds even exceeding the escape speeds of the Galilean moons. 
Given that the surface properties of Amalthea and Thebe are likely comparable  (dirty ice), we assume 
similar ejection speeds also for these small moons.

In our calculations
we shall use some intermediate values of dust grains ejected velocities (a few $\mathrm{km\,s^{-1}}$) which are in the
range of the escape speeds from the Galilean moons. 
Previously, while discussing dust dynamics in the gossamer rings and Thebe extension,
the role of a dust grain initial velocity was neglected.
Taking into account the initial
velocity makes it possible to explain the penetration of
some fraction of dust grains ejected from Thebe into the Thebe extension.
There are some peculiarities of the discussed process. The ejected grain
should have a significant initial velocity with the appropriate direction of propagation.
The most favourable situation appears when a given large dust grain is ejected
from the surface of Thebe in the direction of Thebe's instant azimuthal velocity or, more generally,  
the grain at the
moment of ejection has a significant velocity component in the direction of
the instantaneous azimuthal speed of Thebe and the radial velocity directed into the Thebe extension.
The discussed requirements are easily realized if Thebe and a micrometeorite hitting the surface move
towards each other with antiparallel speed vectors. Note that in such case the relative velocity of 
two bodies can exceed $\mathrm{50\,km\,s^{-1}}$.

According to our calculations dust grains with sizes of the order of  several micrometers
ejected with  velocities $V_{\theta} \geq \mathrm{1.5\,km\,s^{-1}}$  penetrate deep
into the Thebe extension. At the same time smaller grains
(fraction of a micrometer)  penetrate into the Thebe extension only at the initial stage after ejection from Thebe. 
Later on their radial localization
shifts more and more into the Thebe ring. That is why tiny grains appear in the Thebe extension
mainly due to some other mechanism. According to \citet{burns2004} 
tiny dust grains are produced in  Jupiter's rings  due
to "sputtering of surrounding plasma and collisions with gravitationally
focused interplanetary micrometeorites". Later on \citet{dikarev2006} -- based
on these ideas -- developed the model of two-stage dust delivery from satellites to planetary rings.
According to this model large (multi-micrometer-sized) dust grains ejected from the
surfaces of the moons supply the rings with tiny dust grains due to collisions between themselves
and with other micrometeorites. Note that the influence of electric charging on the dust grains
dynamics was neglected in this model.

In this paper we argue that some part of large (from a few micrometers up to multi-micrometer sizes)
electrically charged grains ejected from Thebe due to collisions with micrometeorites
penetrate deep into the Thebe extension. Smaller dust
grains appear subsequently due to fragmentation of large grains in the Thebe extension
(caused by sputtering, mutual
collisions and collisions with micrometeorites or by electric disruption).
Note that small electrically charged dust grains ejected from Thebe migrate rather quickly into the Thebe
ring. At the same time we show that similar grains that appear in the Thebe extension due
to fragmentation of large grains remain in the extension for years.

\section{Basic Equations}

In order to discuss the motion of dust grains in the equatorial inner magnetosphere of Jupiter
we use a cylindrical coordinate system  centered at Jupiter. Suppose that the z-axis is directed upwards
while the magnetic field lines of Jupiter are directed downwards. Two other
coordinates are the radial coordinate in the equatorial plane $r$ and the
angular coordinate $\theta$. The
equations of motion for charged dust grains in this
coordinate system were investigated earlier by \citet{horanyi1991}. We present them
in the following form:
\begin{eqnarray}
&&\frac{d^2r}{dt^2}=r\Omega^2+\frac{q}{r^2}(\Omega -\Omega_J
)-\frac{\mu}{r^2}\nonumber \\
&&\frac{d\Omega}{dt}=-\frac{d\ln r}{dt}\left (\frac{q}{r^3}+2\Omega\right ).
\label{1}
\end{eqnarray}

Here $\Omega =d\theta/dt$ is the angular frequency of the  orbiting dust grain,
$\Omega_J \approx \mathrm{0.000176\,s^{-1}}$ is the angular frequency of Jupiter's rotation, 
$q=Z\Omega_{He}R_{\rm Th}^3 m_e/M_d$, $\Omega_{He}$ is the Larmour
frequency of electrons on the surface of Jupiter, $\Omega_{He}\approx \mathrm{8.5\cdot10^7\,s^{-1}}$, $m_e, M_d$ are the masses of
an electron and a dust grain, respectively, $Z=Q_d/e$ is the amount of elementary
charges on a given dust grain, $e$  is the charge of the electron, $Q_d$ is the
charge on a dust grain, $\mu=G M_J$, $G=\mathrm{6.68\cdot10^{-8}\,cm^3\,g^{-1}\,s^{-2}}$ is the gravitational
constant, $M_J \approx \mathrm{1.9\cdot10^{30}\,g}$ is the mass of Jupiter, 
and $R_{\rm Th}=3.11\,\mathrm{R_J}$ is the radial distance of Thebe's orbit from Jupiter.

Equation~1 describes the motion in the equatorial plane of Jupiter of a grain with the electric charge $Q_d$
and mass $M_d$. The magnetic field has the dipole form. It is assumed that the inner magnetosphere
rigidly corotates with the planet. Two forces -- the gravitational force (term  $\mu/r^2$) and the electric force
(term $ q(\Omega -\Omega_J)/r^2$) -- determine the motion of a dust grain. Previously, the dynamics
of a charged grain ejected with zero initial velocity with respect to the moon was considered
(see, e.g. \citet{horanyi1991}). Here  we discuss the more general and more
realistic case that ejected  grains have initial radial and azimuthal velocities (Sections 3 and 5). This is quite
natural because dust grains are ejected from the moon due to micro-explosions caused by micrometeorite
bombardment. It is important that some ejected grains have higher initial azimuthal velocities than the
azimuthal velocity of the moon.  We show that such large dust grains
(with small ratio $Q_d/M_d$) have equilibrium orbits shifted outwards with respect to the orbit of the moon
(see Figs. 1 and 3).

Note that the electric charge on a dust grain $Q_d$ is not constant. Usually
dust grains lying on the surface of the moon (Thebe in our case) have electric
charges much smaller than the electric charges in equilibrium in space \citep{borisov2006}. That is why after
ejection from the surface dust grains continue to acquire electric charge
approaching an equilibrium value $Q_{\rm st}$.

Let us proceed with the second of Eq.~(1).
It is convenient to seek for $\Omega(t)$ in the form
\begin{equation}
\Omega(t)=\Omega_1(t)\frac{R_{\rm Th}^2}{r^2} .\label{2}
\end{equation}
 The equation for $\Omega_1$ takes the form
\begin{equation}
\frac{d\Omega_1}{dt}=-\frac{q(t)}{R_{\rm Th}^2r^2}\frac{dr}{dt}. \label{3}
\end{equation}

 This equation  can also be  written in the equivalent form which
 is more convenient for an approximate solution in case of a changing charge
 $q(t)$:
 \begin{equation}
 \frac{d\Omega_1}{dt}=\frac{1}{R_{\rm Th}^2}\frac{d}{dt}\left (\frac{q}{r}\right
 )-\frac{1}{r R_{\rm Th}^2}\frac{dq}{dt}. \label{4}
 \end{equation}

If a dust grain starts its motion from the surface  of Thebe with negligible
velocity, the initial condition for $\Omega_1$ is $\Omega_1(0)=\Omega_{\rm Th}$, where $\Omega_{\rm Th}\approx \mathrm{0.000107\,s^{-1}}$ is the
angular orbital rotation frequency of Thebe. If the initial azimuthal velocity
 $V_{\theta}$ in the system moving with Thebe is finite $\Omega_1(0)=\Omega_{\rm Th} +\Delta\Omega$, where $\Delta \Omega
=V_{\theta}/R_{\rm Th}.$  As a
 result the frequency $\Omega$ takes the form
\begin{equation}
\Omega(t)=(\Omega_{\rm Th}+\Delta\Omega)\frac{R_{\rm Th}^2}{r^2}+\frac{1}{r^2}\left [\frac{q(t)}{r}\right ]_0^t-
\frac{1}{r^2}\int_0^t\frac{1}{r}\frac{dq}{dt}\,dt .
\label{5}
\end{equation}

Here $q(0)=Z_0\Omega_{He}R_{\rm Th}^3 m_e/M_d$, $Z_0$ is the amount of electrons
 on a given dust grain lying on the surface of the moon $Z_0=Q_0/e$,
$r_0=R_{\rm Th}$ is the initial radial coordinate.

If the electric charge is constant, $q=q_{\rm st}$, the derivative is $dq/dt=0$.
In this case we find
\begin{equation}
\Omega(t)=\Omega_0\frac{R_{\rm Th}^2}{r^2}+\frac{q_{\rm st}}{r^2}\left (
\frac{1}{r}-\frac{1}{R_{\rm Th}}\right ), \label{6}
\end{equation}
where $\Omega_0$ is the initial angular frequency.
If the radial coordinate of a given dust grain $r(t)$ changes more significantly than the electric charge $q(t)$
($|\frac{dr}{rdt}|>>|\frac{dq}{qdt}|$) then we receive an approximate solution which is more general than the result
given in Eq.(6):
\begin{equation}
\Omega(t)=\Omega_0\frac{R_{\rm Th}^2}{r^2}+\frac{q(t)}{r^2}\left (
\frac{1}{r}-\frac{1}{R_{\rm Th}}\right ). \label{7}
\end{equation}

Here $\Omega_0=\Omega_{\rm Th}+\Delta\Omega$. With  increasing time $q(t)\to
q_{\rm st}$, where $q_{\rm st}$ is the  equilibrium charge on a dust grain in space.
It follows from Eqs.~(5) and (7) that for the moment $t=0$ the frequency $\Omega=\Omega_{\rm Th}+\Delta\Omega$, as it should
be. Note that the sign of $q$ is positive for negatively charged grains
because the magnetic field of Jupiter is directed downwards. 

It is easy to check that the expression for $\Omega(t)$ given
by Eq.~(5) is the solution of the second of Eq.~(1). As a result we
may conclude the following. First,
the last term on the right-hand side of Eq.~(7) is positive for radial
distances $r$ less than the orbit of Thebe $R_{\rm Th}$. Second, the smaller the
variations  $|R_{\rm Th} -r|$, the less significant is the influence of the electric
charge on the angular frequency. For a rather large time $t\geq t_{\rm st}$ when $Q(t)=Q_{\rm st}$
we may designate $\Omega_0 R_{\rm Th}^2- q_{\rm st}/R_{\rm Th}=J$ and
express the angular frequency as
\begin{equation}
\Omega(t)=\frac{J}{r^2}+\frac{q_{\rm st}}{r^3} .\label{8}
\end{equation}

It can be shown that our Eq.~(8) coincides with Eq.~(3b) of \citet{horanyi1991}.
But for a smaller period of time $t\leq t_{\rm st}$ we may use a more general expression:
\begin{equation}
\Omega(t) =\frac{J}{r^2} +\frac{q(t)}{r^3} -\frac{q(t)-q_{\rm st}}{r^2R_{\rm Th}} .\label{9}
\end{equation}

The frequency $\Omega(t)$ given by Eq.~(9) can be substituted into the
first Eq.~(1) to obtain an equation describing the radial motion of a
dust grain. Assuming as in the paper by \citet{horanyi1991} that the terms with the electric charge are small
and retaining only the terms of the first order with respect to $q$, we arrive at an
 approximate equation:
\begin{equation}
\frac{d^2 r}{dt^2}=\frac{J^2}{r^3}+\frac{3
qJ}{r^4}-\frac{2(q-q_{\rm st})J}{r^3R_{\rm Th}}-\frac{\mu+q\Omega_J}{r^2} .\label{10}
\end{equation}

\section{Neutral Dust Grain Dynamics}

Let us introduce the dimensionless coordinate $\rho=r/R_{\rm Th}$ instead of the 
radial distance $r$.
If the electric charge is  negligible ($q(t)\to 0$) Eq.~(11) describes the
radial motion of a neutral dust grain:
\begin{equation}
\frac{d^2\rho}{dt^2}=\frac{J^2}{R_{\rm Th}^4\rho^3}-\frac{\mu}{R_{\rm Th}^3\rho^2}.
\label{11}
\end{equation}

It can be  shown with the help of Eq.~(11) that
at the moment $t=0$ when $\rho =1$ the radial force is positive
if $\Delta\Omega >0$. With increasing $\rho$ this force decreases and
at $\rho=\rho_{*}$, where $\rho_{*}=J^2 (R_{\rm Th}\mu)^{-1}$ it
becomes equal to zero. For larger distances $\rho >\rho_{*}$ this force is
negative and it pulls dust grains  radially inwards into the Thebe ring.

Let us discuss the radial motion of a neutral dust grain analytically. It
follows from Eq.~(11):
\begin{equation}
\left (\frac{d\rho}{dt}\right
)^2=\frac{2\mu}{R_{\rm Th}^3\rho}-\frac{\Omega_0^2}{\rho^2}+W_0 ,\label{12}
\end{equation}
where $W_0$ is added on the right-hand side of Eq.~(12) to include the correct value of the initial radial
velocity $V_r$
\begin{equation}
W_0=\Omega_0^2-\frac{2\mu}{R_{\rm Th}^3}+\frac{V_r^2}{R_{\rm Th}^2}. \label{13}
\end{equation}

From Eq.~(13) we find the maximal radial excursion $\rho_{\rm max}$ for a given dust grain:
\begin{equation}
\rho_{\rm max}=-\frac{\mu}{R_{\rm Th}^3W_0}+\sqrt{\left
(\frac{\mu}{R_{\rm Th}^3W_0}\right )^2+\frac{\Omega_0^2}{W_0}} .\label{14}
\end{equation}

\begin{figure}[tb]
		\includegraphics[width=0.8\textwidth]{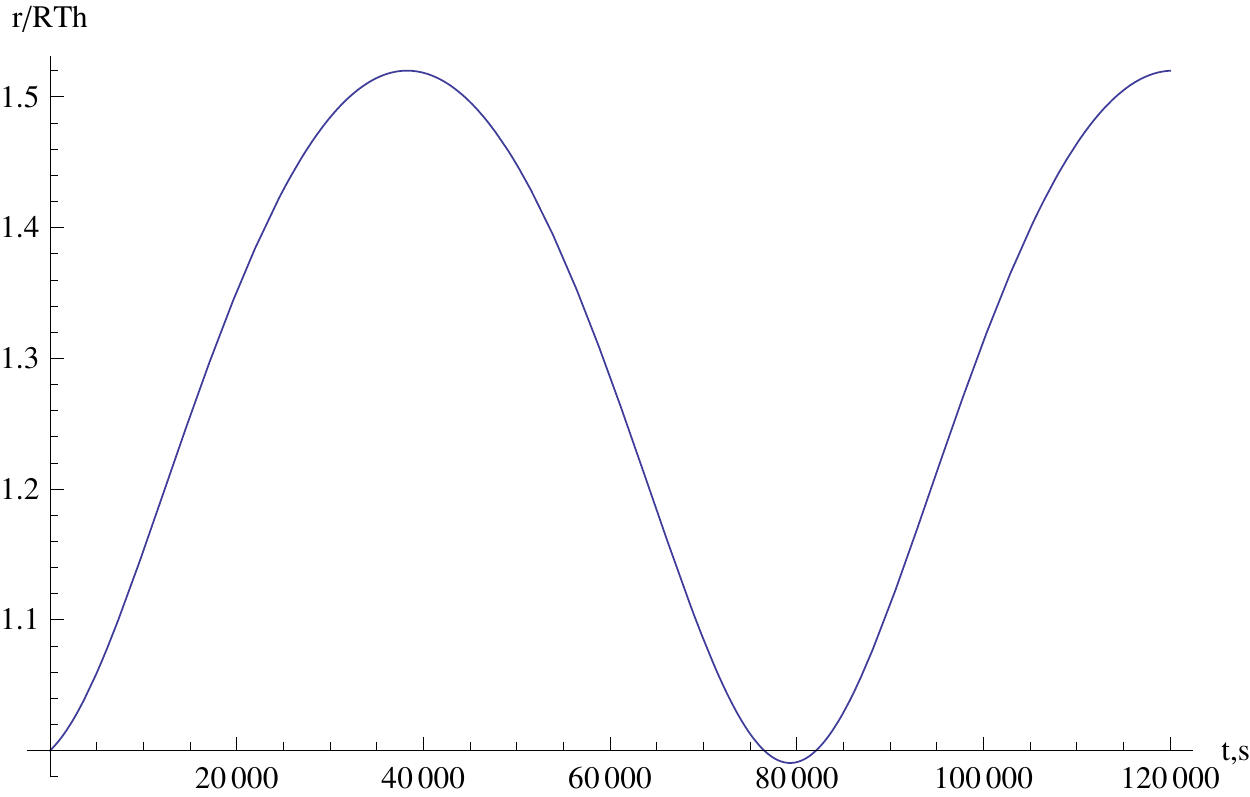}
	\caption{Radial motion of a dust grain without electric charge. The
horizontal axis gives the time in seconds  and  the vertical axis the relative radial
coordinate $r/R_{\rm Th}$. The initial radial velocity directed into the Thebe extension is $V_r(t=0)=\mathrm{1.5\,km\,s^{-1}}$.
The initial azimuthal velocity with respect to the azimuthal velocity of Thebe is  $V_{\theta}=\mathrm{2.2\,km\,s^{-1}}$. The
grain starts its motion at $t=0$ from the orbit of Thebe.}
	\label{fig01}
\end{figure}

In Figure~\ref{fig01} we show the radial
distance of a dust grain 
for an initial azimuthal $V_{\theta}=\mathrm{2.2\,km\,s^{-1}}$ and radial
$V_r=1.5\,\mathrm{km\,s^{-1}}$ velocities. The initial azimuthal velocity is taken in the direction of Thebe's orbital motion, 
and the radial
velocity is directed towards the Thebe extension. It can be seen that neutral dust grains with 
such significant initial velocities propagate rather far outside of the Thebe orbit
regardless of their size. In our case, the grain propagates up to
$\rho_{\rm max}=1.52$, which is equivalent to approximately 115,000~km. According to
Fig.~\ref{fig01} it takes a significant amount of time $\Delta t \approx 22$~hours for a neutral dust
grain to penetrate into the Thebe extension and to return to its initial
radius $\rho =1$. The maximum radial velocity reaches the value
$V_{\rm max}\approx \mathrm{6\,km\,s^{-1}}$. Note that such dynamics is valid for all neutral dust grains
(with given initial velocities) regardless of their mass. The situation
changes if we take into account that dust grains are negatively charged and
the ratio of the electric charge to mass $Q_d/M_d$ varies significantly for
grains with different sizes $a$. Indeed,  under equilibrium conditions the electric charge on a 
given dust grain with  radius $a$
is $Q_d=\phi_{\rm st} a$, while the mass of such grain is $M_d=(4\pi/3) \rho_{\rm dust} a^3$, where $\phi_{\rm st}$ is the electric potential in the
medium, $\rho_{\rm dust}$ is the dust density. The ratio $Q_d/M_d$ is inversely
proportional to $a^2$. Thus, the electric force influences more significantly
the motion of small dust grains. That is why the motion of neutral dust
grains discussed in this section can be considered as a good approximation
for the dynamics of large multi-micrometer sized dust grains.  We will discuss the influence of the 
electric charge and mass
on the dust grains dynamics in more  detail in the next sections.

\section{Charging of a Dust Grain on the Surface of the Moon and in Space}

In this section we discuss charging of dust grains in the inner magnetosphere of Jupiter taking into account
the Divine model \citep{divine1983}. At the beginning we estimate the
initial electric charge on a grain lying on the surface of the moon (Thebe). Then we
consider the equilibrium charge on the same grain ejected from the surface. At
the end we investigate how the electric charge on a given grain grows in time
after ejection from the moon.

The problem of dust grains charging was discussed in many papers,
see, e.g.,\citet{whipple1981,goertz1989,whipple1985,havnes1987,horanyi1996b,borisov2006}.
We present some estimates for the typical
electric charge on a dust grain with  radius $a$ lying on the surface of the
moon and also in space after ejection.

The process of dust grain charging is considered in the system of coordinates rotated with the magnetic field.
This means that there is no corotating electric field acting on electrons and ions but instead there
exists an azimuthal  velocity component of a dust grain with respect to the plasma $V_0=R_d(\Omega_d-\Omega_J)$,
where $R_d$ is the radial distance of a dust grain from Jupiter,
$\Omega_d$ is its angular velocity, $\Omega_J$ is the angular velocity of the planet (and the magnetic field).
In  general this velocity  should be taken into account when discussing dust grain charging by ions.
Below  it is shown that in our case the additional flux of ions  associated with
this velocity is smaller than the  well-known flux of ions in the vicinity of a grain
(see, e.g. \citet{whipple1981,goertz1989,whipple1985,havnes1987,horanyi1996b}) that
provides its charging. Hence, to a first approximation the velocity $V_0$ can be
neglected.

In order to calculate the
equilibrium potential we assume that far away from the moon electrons have a Boltzmann distribution
with the temperature $T_e$ and  concentration $N_e$:
\begin{equation}
F_e=\frac{N_e}{\pi^{3/2}V_{T_e}^3}\exp\left
(-\frac{v^2}{V_{T_e}^2}\right ), \label{15}
\end{equation}
where $V_{T_e} =\sqrt{2T_e/m_e}$ is the thermal velocity, and
$m_e$ is the mass of an electron. Ions are assumed to have a similar
distribution:
\begin{equation}
F_i=\frac{N_i}{\pi^{3/2}V_{T_i}^3}\exp\left
(-\frac{v^2}{V_{T_i}^2}\right ), \label{16}
\end{equation}
where $N_i=N_e=N_0$ is the plasma concentration, $T_i$
is the temperature and $V_{T_i}=\sqrt{2T_i/M_i}$ is the thermal velocity of
ions, $M_i$ is the mass of ions.

Our aim is to estimate the electric potential $\phi$ on the surface of the moon in a plasma that contains
electrons and ions with the distribution functions (15), (16) far away from the surface. This potential
should be negative because the thermal speed of electrons is much higher than
the thermal speed of ions $V_{Te}>>V_{Ti}$. Let us introduce the Debye radius
$R_D=V_{Te}/\Omega_{Pe}$. Here  $\Omega_{Pe}=(4\pi e^2N_e/m_e)^{1/2}$
is the plasma frequency, $e$ is the charge of an electron. The Debye radius in the
vicinity of Thebe according to the Divine model is
of the order of $R_d\approx 1-2$ m.

Electrons are retarded by the
negative potential. Their concentration in the plasma
with the potential $\phi$ is  $N_e=N_0\exp
(e\phi/T_e)$ while the velocity distribution is still determined by Eq.~(15).
The flux of electrons on to the negatively charged surface according to \citet{havnes1987} is:
\begin{equation}
\Phi_e=\sqrt{\frac{2}{\pi}} V_{Te} N_0\exp\left (\frac{e\phi}{T_e}\right ).
\label{17}
\end{equation}
At the same time the flux of ions on to the same surface is:
\begin{equation}
\Phi_i=\sqrt{\frac{2}{\pi}}V_{Ti} N_0 \left (1-\frac{e\phi}{T_i}\right )\label{18}.
\end{equation}
Equating the fluxes of electrons and ions we find for the electric potential
under equilibrium conditions and equal temperatures $T_e=T_i$:
\begin{equation}
(1+F_{st}(a))\exp(F_{st}(a))=\frac{V_{Te}}{V_{Ti}} \label{19},
\end{equation}
where $F_{st}=-e\phi_{st}/T_e$.  For equal temperatures of electrons and ions and heavy ions
($M_i/m_e\approx 3\cdot10^4$) the electric potential on the surface is
$e\phi_s/T_e\approx -3.6$.
The electric field above the surface of the moon is described by the
Poisson equation:
\begin{equation}
\frac{d^2\phi}{d z^2}=4\pi e(N_e(\phi)-N_i(\phi)), \label{20}
\end{equation}
where $z$ is the coordinate across the surface of the moon.
A detailed calculation of the surface electric field can be found in \citet{borisov2006}.
According to this paper the characteristic length that determines the electric
field is $L_z\approx (3-4) R_D$. So we estimate the vertical surface electric field as
$E_s\approx (0.3\div 0.4)\,\mathrm{V\,cm^{-1}}$. This field
is connected with the density $\sigma$ of the electric charge on the surface that provides
such electric field $E_s$, see \citet{borisov2006}. Indeed, after integration of Eq.~(20) across the
surface we arrive at the relation
\begin{equation}
\frac{d\phi}{dz}|_{+0}-\frac{d\phi}{dz}|_{-0}=4\pi \sigma. \label{21}
\end{equation}
This equation allows us to estimate the charge on a dust grain with the radius a lying on
the flat surface. Taking into account that the cross-section of a dust grain
is $\Sigma =\pi a^2$ we find the electric charge on a  grain lying on the
surface as $Q_d=0.5 a^2 E_s$. For example, the grain with the radius $a=5~\mu\mathrm{m}$
lying on the surface of the moon (Thebe) contains 5 to 6 electrons ($Z=5-6$).

After ejection from the moon a dust grain acquires additional negative electric
charges until its potential $\phi_{st}$ reaches an equilibrium value (which is achieved when the
flux of electrons hitting a given grain becomes equal to the flux of ions). It
is more convenient to use the dimensionless
parameter $F=-e\phi/T_e$ instead of the potential $\phi$.

It is interesting that the electric potential determined above by Eq.~(19) in the
vicinity of a grain can accelerate  ions up to the velocity $V_b\approx
(-2e\phi_{st}/M_i)^{1/2}$ towards a grain.
This velocity comprises $V_b\approx 35\,\mathrm{km\,s^{-1}}$  and it is higher than the thermal
speed of ions ($V_b\approx 1.9\,V_{Ti}$). At the same time the
azimuthal velocity of a dust grain with respect to the magnetic field
(mentioned at the beginning of this section) in the
vicinity of Thebe is $V_0\leq  15\,\mathrm{km\,s^{-1}}$. As a result the flux of ions given by
Eq.~(18) is significantly larger than the flux $\Phi_0$ associated with the azimuthal velocity
$V_0$ which provides additional charging mainly from one side $\Phi_0\approx \pi a^2
N_0 V_0$.

Note that while calculating the electric potential
we have not included  the flux of  secondary electrons. This
flux contains several parameters whose values are rather uncertain. For
the same values of parameters that were used earlier in relation with the inner
magnetosphere of Jupiter \citep{horanyi1991} we find that the deviation
of the surface potential from the one estimated above is only approximately $1.5\,\%$.
Taking into account that the electric potential
$\phi_{st}$ is connected with the equilibrium electric charge on a given dust grain by the 
relation $\phi_{st}=Q_d/a$, we are able to estimate the equilibrium electric charge on a
dust grain with  radius a in space. This calculation shows that in the vicinity of
Thebe a dust grain with  radius a 
should have the charge $Q_d\approx 3.6\,aT_e/e$ under equilibrium conditions. For a 
grain with  radius
$a=5~\mu\mathrm{m}$ the charge is $Q_d\approx 5\cdot 10^5 e$ ($Z=5\cdot10^5$).
It should be mentioned that the electric charge on a grain lying on the surface
is proportional to the square of its radius while in space the equilibrium
charge is proportional only to its radius. Nevertheless for $a=5~\mu\mathrm{m}$ the
charge in space is approximately five orders of magnitude higher than the
charge on the same grain lying on the surface.

Now we need to investigate how the electric charge on a given dust grain changes
in time. For this purpose we take into account the difference of the fluxes
of electrons and ions that provide the growth of the electric charge
\begin{equation}
\frac{dQ_d}{dt}=4\pi a^2 e(\Phi_i(a)-\Phi_e(a)) \label{22}.
\end{equation}
This equation rewritten in terms of the dimensionless function $F$ takes the form
\begin{equation}
\frac{dF(a)}{dt}=\frac{\sqrt{2\pi}V_{Te}a}{R_d^2}\left
(\exp (-F(a))-\frac{V_{Ti}}{V_{Te}}(1+F(a))\right ). \label{23}
\end{equation}
The dimensionless potential $F(a)$ grows in time approaching its equilibrium value
$F_{st}(a)\approx 3.6$. With the help of Eq.~(24) it is possible to estimate the
characteristic period $\Delta t$ of charging:
\begin{equation}
\Delta t \approx \frac{R_d^2}{\sqrt{2\pi}V_{Te}a} \label{24}.
\end{equation}
It follows form Eq.~(24) that the period $\Delta t$ is inversely proportional to the radius of a grain.
For a dust grain with  radius $a=1~\mu\mathrm{m}$ this time is of the order of one second.
Note that $\Delta t$ describes the characteristic period for the initial
stage of charging. More accurately, the variation of the potential $F(a)$ can
be obtained solving  Eq.~(24) numerically. In Fig.~\ref{fig03} we show the growth in time of the
potential $F(a)$ for the grain size $a=1~\mu\mathrm{m}$. On the horizontal axis the dimensionless time $t_1=t/\Delta t$
is given. It is seen that the process of charging
becomes  slower and slower with time. It takes the time $15 \Delta t$
to achieve $0.7F_{st}(a)$.

\section{Dust Grain Dynamics in the Thebe Extension}

\begin{figure}[tb]
		\includegraphics[width=0.8\textwidth]{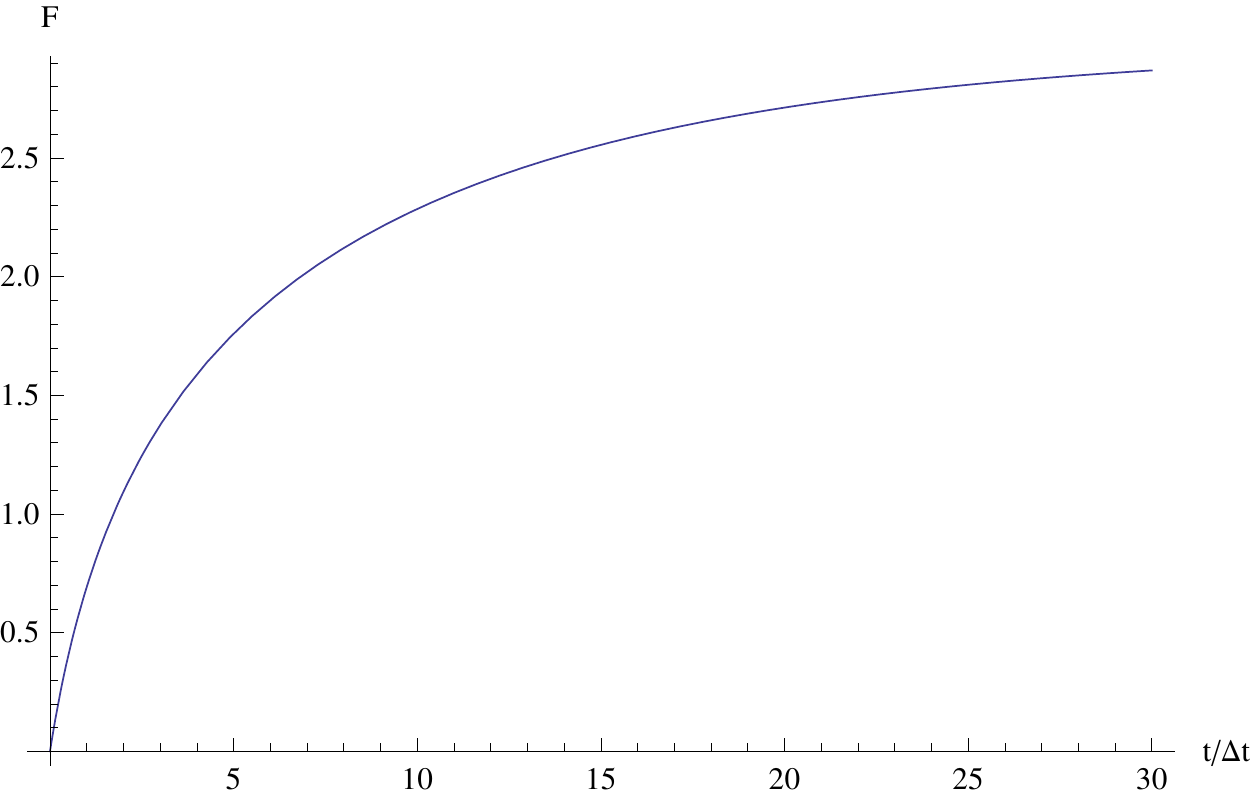}
	\caption{Growth of the electric potential towards its  equilibrium value in time on a given dust grain
with the radius $a=1\,\mathrm{\mu m}$ after its ejection from Thebe at $t=0$. On the
horizontal axis time in dimensionless units $t/\Delta t$ is presented.}
	\label{fig03}
\end{figure}

The dust grain dynamics after ejection from the moon with some finite initial
velocity can be investigated numerically. The radial displacement in time of the charged dust grains with different sizes
$a=5~\mu\mathrm{m}$  and $a=0.3~\mu\mathrm{m}$  are calculated taking into account the equations of motion (see Section 2).
The results are shown in Figs.~\ref{fig04} and ~\ref{fig05}. The initial
azimuthal and radial velocities are $V_{\theta}=2.2\,\mathrm{km\,s^{-1}}$ and
$V_r=1.5\,\mathrm{km\,s^{-1}}$ (the same for both grains). The calculations are carried out based on the Divine
model. It is seen that the heavy charged dust grain penetrates
rather deep into the Thebe extension (approximately 100,000 km). This result
qualitatively corresponds to the motion of a neutral dust grain (see Fig.~\ref{fig01}).
Only the radial excursion into the Thebe extension is somewhat smaller due to the action of the electric force. At the
same time the submicrometer dust grain at the beginning slightly penetrates into the Thebe
extension and later on its radial motion shifts into the Thebe ring.  It is
interesting to mention that the period of radial oscillations for a smaller dust
grain is much shorter than that for the larger one. Such an effect was not
discussed before. It happens because strong electric charging of a dust grain
(that exists due to high plasma
concentration and temperature) influences the  period of oscillations.

\begin{figure}[tb]
		\includegraphics[width=0.8\textwidth]{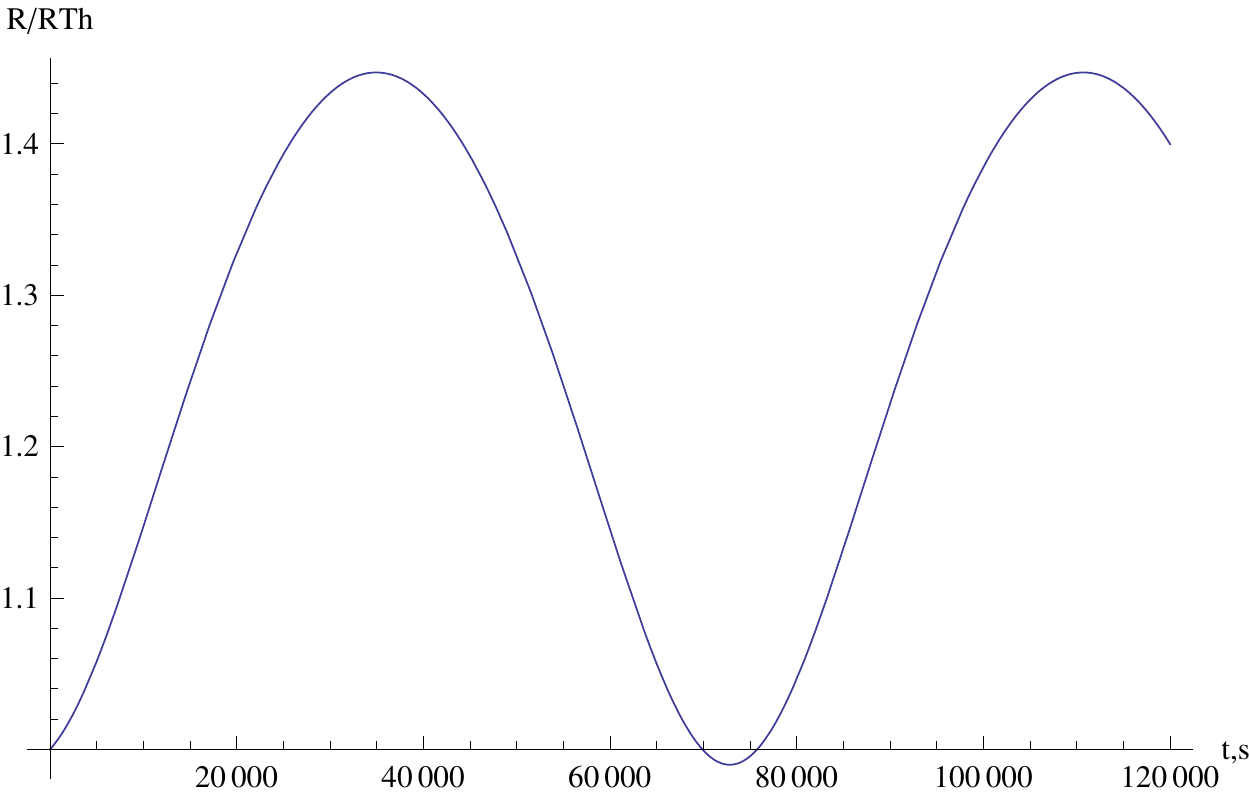}
	\caption{Radial oscillations in time of a charged dust grain with the radius
$a=5\,\mathrm{\mu m}$  ejected from Thebe. The horizontal axis shows the time in seconds and  the
vertical axis the radial coordinate in dimensionless units $r/R_{\rm Th}$. The initial radial velocity 
directed outwards is
$V_r=\mathrm{1.5\,km\,s^{-1}}$ and the azimuthal velocity with respect to the azimuthal
velocity of Thebe is $V_{\theta}=\mathrm{2.2\,km\,s^{-1}}$.}
	\label{fig04}
\end{figure}

\begin{figure}[tbh]
		\includegraphics[width=0.79\textwidth]{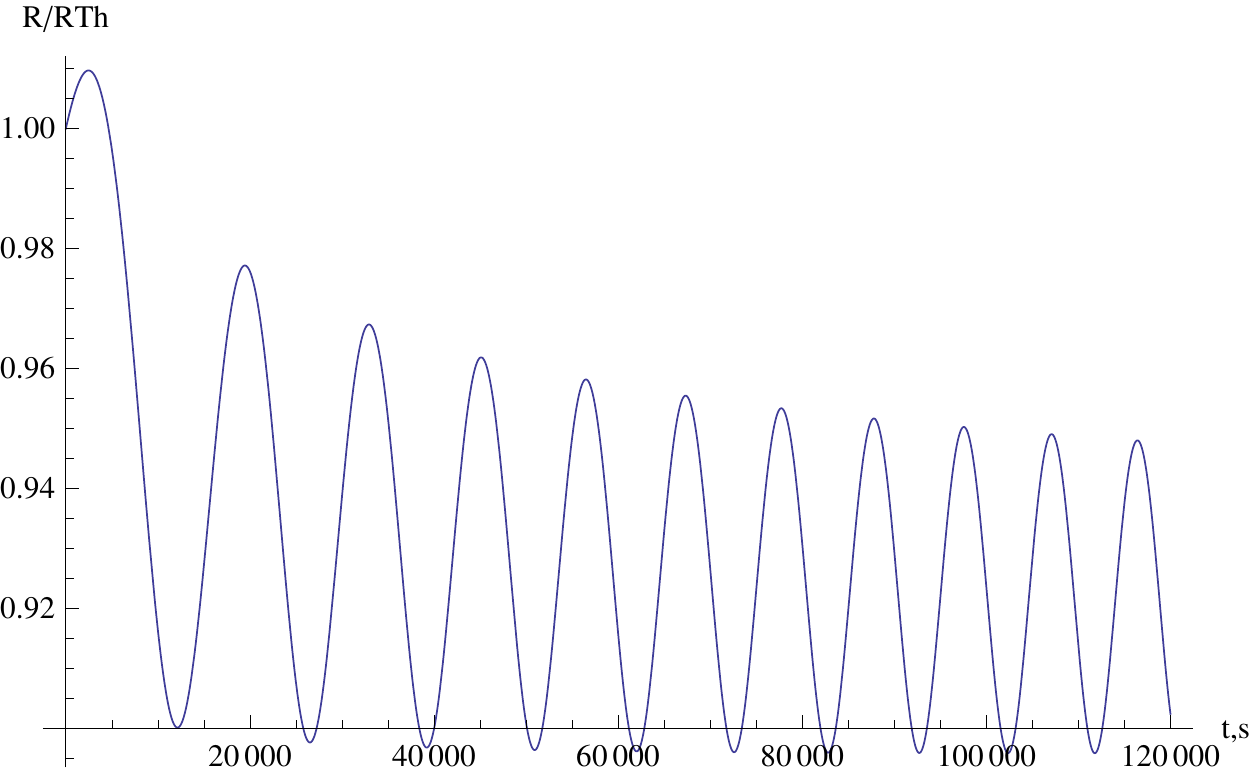}
	\caption{The same as in Fig.~\ref{fig04}, except that the radius of a dust grain is
 $a=0.3\,\mathrm{\mu m}$.}
	\label{fig05}
\end{figure}

\begin{figure}[h]
		\includegraphics[width=0.79\textwidth]{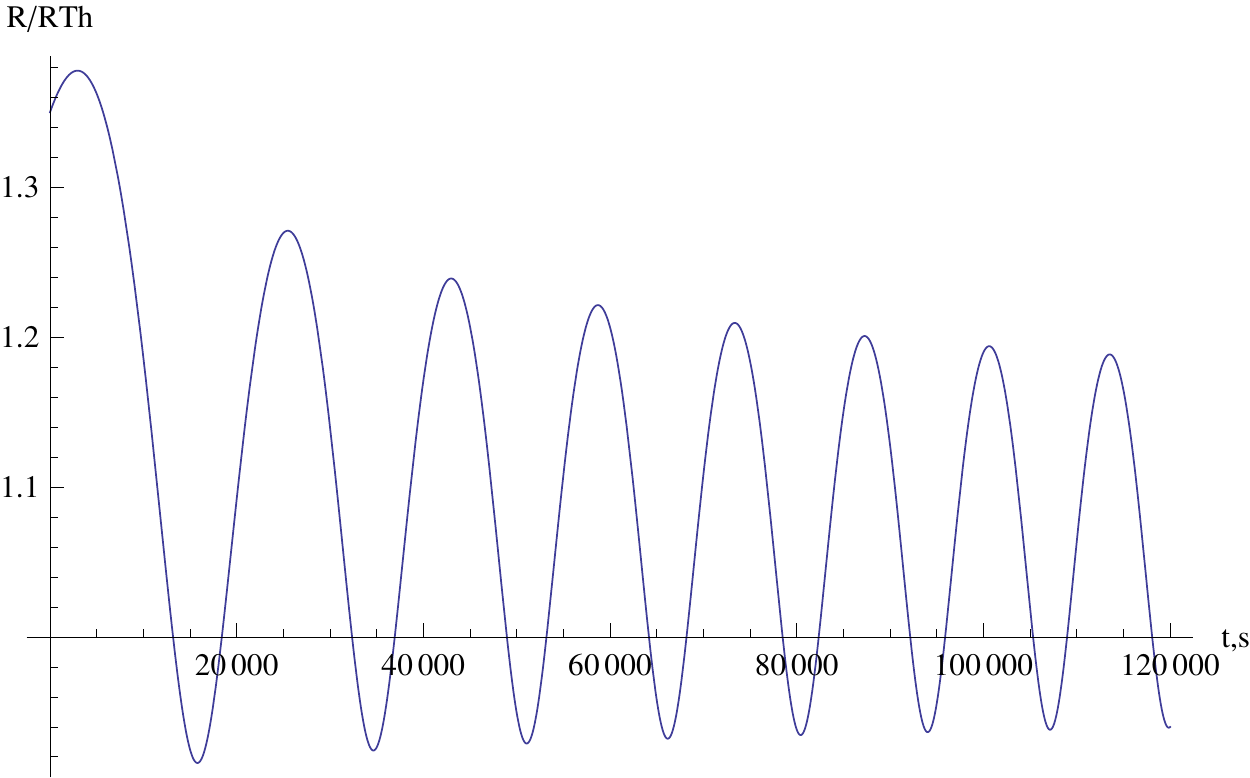}
	\caption{Radial oscillations of a small dust grain
($a=0.3\,\mathrm{\mu m}$) ejected  at  $r_0=1.22\,R_{\rm Th}$ with the
initial radial velocity equal to the corresponding velocity
of a large dust grain $V_r(t=0)=2.8\,\mathrm{km\,s^{-1}}$. The initial azimuthal frequency
$\Omega(t=0)=0.000078\,\mathrm{s}^{-1}$  coincides with the angular frequency of a large
dust grain located at  $r_0=1.22\,R_{\rm Th}$.}
	\label{fig06}
\end{figure}

All calculations above were carried out assuming that dust grains (small and large)
were ejected from the surface of Thebe. Now we suppose that a small dust grain
appears in the Thebe extension due to  fragmentation of a larger dust grain (e.g. in
collision with a micrometeoroid). In Fig.~\ref{fig06}  we present one such example. A dust
grain with the size $a=0.3\,\mathrm{\mu m}$ is released at t=0 from a large grain in the vicinity
of $r=1.22 R_{\rm Th}$. It is assumed that the small grain is ejected
with the radial velocity $V_r(t=0)=\mathrm{2.8\,km\,s^{-1}}$ and the angular frequency  $\Omega(t=0)=\mathrm{0.000078\,s^{-1}}$.

In contrast to the results presented in Fig.~\ref{fig05} this small
dust grain does not migrate into the Thebe ring but continues to oscillate
in the Thebe extension. In the equilibrium state the oscillations take place between $r_1 \approx 0.96 R_{\rm Th}$ and $r_2\leq
1.22 R_{\rm Th}$. This is a new result because earlier it
was found that the stationary orbit of dust grains should be shifted inwards
from the moon due to the action of the electric field, see, e.g. \citet{horanyi1991}.
Such distinction appears because of the different initial conditions.

To better understand  the situation, we have calculated
the potential describing the radial motion of a small dust  grain with a constant
electric charge:
\begin{equation}
\frac{1}{2}\left (\frac{d\xi}{dt}\right )^2=U(\xi) +W_0. \label{44}
\end{equation}
\begin{figure}[h]
		\includegraphics[width=0.75\textwidth]{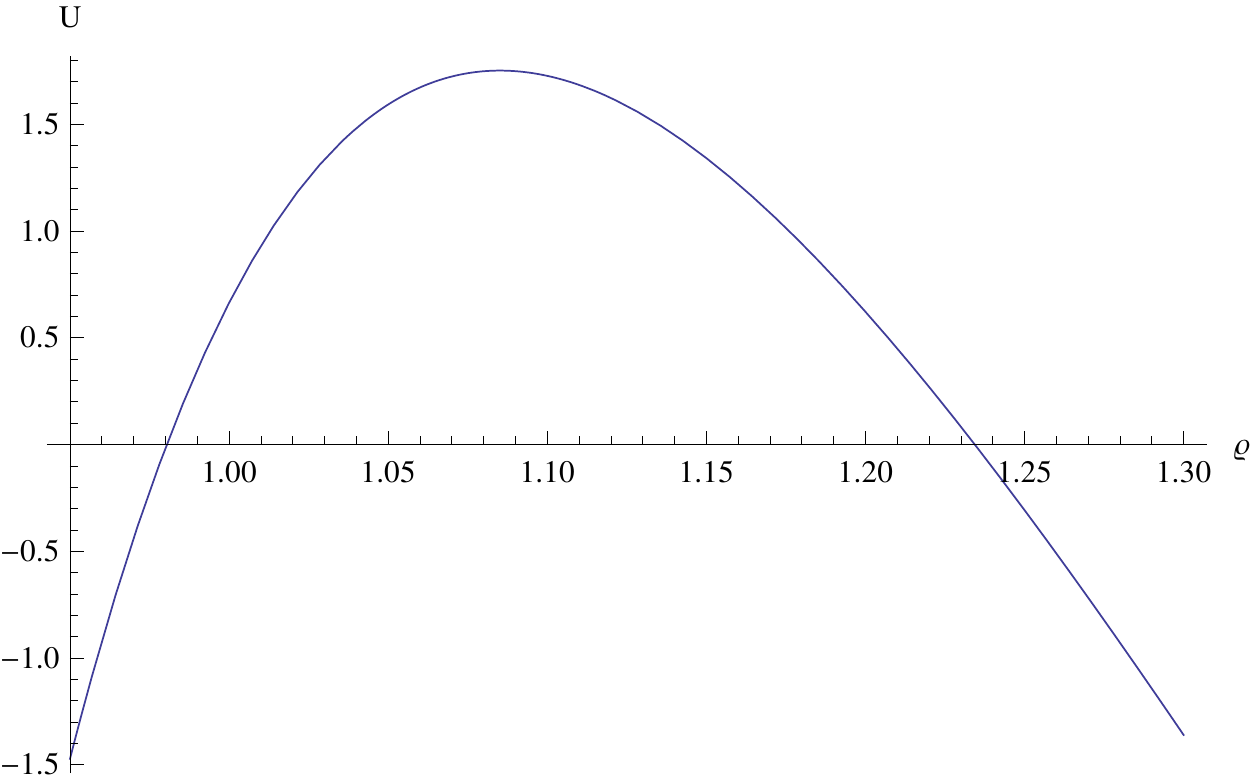}
	\caption{Distribution in space of the potential $U$ for a
small dust grain ($a=0.3\,\mathrm{\mu m}$) with an equilibrium electric charge.
 The initial velocities are the same as in Fig.~\ref{fig06}. On the horizontal axis
the dimensionless coordinate $r/R_{\rm Th}$ is given. The region with $U>0$
corresponds to the range of radii where dust the grain oscillates.}
	\label{fig07}
\end{figure}
Here $U(\xi)$ is the potential
\begin{eqnarray}
U(\xi)=-\frac{\Omega_0^2\xi_0^4}{2\xi^2}-\frac{\Omega_0q}{R_{\rm Th}^3\xi^2}\left
(\frac{1}{\xi}-\frac{1}{\xi_0}\right
)-\frac{q^2}{2R_{\rm Th}^6\xi^2}\left
(\frac{1}{\xi}-\frac{1}{\xi_0}\right )^2+\frac{\mu+q\Omega_J}{R_{\rm Th}^3\xi}, \label{45}
\end{eqnarray}
where $\xi=r/R_{\rm Th}$, $\Omega_0$ is the initial angular frequency $\Omega_0=\Omega(t=0)$,
$\xi_0$ is the initial radial position of a given dust grain,
$W_0$ is a constant added in Eq.~(25) to provide the correct initial velocity. 

In
Fig.~\ref{fig07}  we present the potential $U_1(\xi)=U(\xi)+ W_0$ for a dust grain starting from
$\xi_0=1.22$  with the angular frequency $\Omega_0=\mathrm{0.000078\,s^{-1}}$.
The dust grain oscillates between $r_1\approx
0.96 R_{\rm Th}$ and $r_2 \approx 1.22 R_{\rm Th}$ which corresponds to the results of the 
numerical computations presented in Fig.~\ref{fig06}.

\section{Discussion and Conclusion}

We have suggested a mechanism  for the filling of the Thebe
extension by dust grains. Our calculations are based on the Divine model
which assumes that thermal plasma in the inner magnetosphere of Jupiter is
rather warm and dense, i.e. at the orbit of Thebe $T_e=45~\mathrm{eV}$ and
$N_e=50\,\mathrm{cm^{-3}}$. As a result  the equilibrium electric potential on  dust
grains in the Thebe extension should be high, $\phi \approx -162~\rm V$, and
the total electric charge on a micrometer-size  grain is
of the order of $Q_d\approx 10^5 e$. Due to such high values of the
potential and the electric charge, dust grains orbiting  Jupiter are almost
insensitive to the solar UV radiation at the sunlit side of the planet. Hence, the mechanism
of shadow resonances as suggested by \citet{hamilton2008} does not work under these conditions.

According to our analysis the dynamics of strongly charged
small dust grains (fraction of a micrometer) in the vicinity of Thebe's orbit can be
quite different depending on the initial conditions. Small dust grains ejected
from Thebe even with  rather high radial velocity only slightly penetrate
into the Thebe extension and later on  gradually migrate into the Thebe ring. Examples of radial oscillations
for grains with different masses are presented in Figs.~\ref{fig04} and ~\ref{fig05}. At the same time,
negatively charged small  grains starting their motion in the Thebe extension remain
there for a long time oscillating between two turning points $r_1$ and
$r_2$. The corresponding example is presented in Fig.~\ref{fig06}. The positions of
these turning points depend on the initial conditions with which a given
small grain starts its motion (radial position and initial velocities). For
example, if a small grain ($a=0.3\,\mathrm{\mu m}$) starts from a large
dust grain  due to fragmentation at $r_0=1.22 R_{\rm Th}$ with the radial velocity and the azimuthal frequency which
coincide with the corresponding parameters of a large grain ($V_r=\mathrm{2.8\,km\,s^{-1}}$,
$\Omega_0=\mathrm{0.000078\,s^{-1}}$), it oscillates in the equilibrium case
between $r_1\approx 0.96~R_{\rm Th}$ and $r_2\approx 1.22~R_{\rm Th}$. This means that
such grain penetrates rather deep into the Thebe extension (up to approximately 50,000\,km). 

Our results have a quite clear physical meaning.
Assume that an additional moon U orbits Jupiter at some distance $R_U > R_{Th}$.
At the moment $t=0$ a small dust grain is ejected from this moon and begins to move with
some negative electric charge. It was shown by \citet{horanyi1991} that negatively charged
 grains ejected outside the synchronous orbit of Jupiter oscillate at  radial
distances slightly less than the orbit of the moon. In our case a large dust grain plays
the role of an additional moon. As shown in Fig.~5, if the angular velocity
of a small dust grain coincides with the that of a large grain (the moon U), after some time
 it oscillates at  radial distances slightly less than the instant radial
distance of the large grain at t=0. Only at the beginning due to its finite initial radial velocity, the
small  grain makes somewhat larger radial excursions. As for large grains they
are weakly influenced by the corotating electric field due to the small ratio of their 
electric charge to mass $Q_d/M_d$.
At the same time it is important to note that we have  considered the case when
large dust grains were ejected from Thebe with a non-zero velocity component parallel to the
angular velocity of Thebe. This additional velocity stems from the ejection
of the particles from Thebe's surface. 
Due to this,  equilibrium orbits for such grains are shifted
towards larger radial distances which is supported by numerical modelling (see Figs. 1 and 3)}.

As mentioned before, the equilibrium electric charge on a  dust
grain in the Thebe extension should be high. Hence, the local electric field in the vicinity
of such a grain is also high. After ejection of a grain from the surface the electric
charge and the electric field grow in magnitude, approaching their equilibrium
values (see Fig.~\ref{fig03}).
If a large grain is porous it might  disintegrate into smaller fragments in case
the electric force becomes strong enough (i.e. the electrostatic stress exceeds the tensile strength),
see, e.g. \citet{stark2015} and \citet{boehnhardt1987}.
Rather crude estimates confirm that the
electric disruption could be efficient in filling the Thebe
extension with small dust grains. Note that such a process should operate also in the gossamer rings.
The other sources of small dust grains in the
Thebe extension are plasma sputtering,  mutual collisions of large grains and their collisions with
micrometeorites. All these processes are slow in comparison with
the electric disruption. Note that due to electric
disruption tiny grains (hundreds and tens of nanometer)  could appear, see
 \citet{stark2015} and \citet{boehnhardt1987}. This means that not only Io  is a source of tiny
grains in the inner magnetosphere of Jupiter (as it was discussed by \citet{gruen1998}) 
 but also electrostatic disruption of rather large negatively charged
dust grains can occur in the vicinity of Thebe's and Amalthea's orbits about 
Jupiter. 

So far, the measurements by the Juno  spacecraft probe the Jovian magnetosphere only down to
 $R\approx 10\,\mathrm{R_J}$, showing that the ion concentrations strongly increase towards 
Jupiter \citep{kim2020}. In the future Juno  will hopefully  provide us with measurements 
of charged particles much closer to Jupiter that will  allow us
to better constrain the physical processes driving  dust particle motion in the Thebe extension.

\section*{Acknowledgments}
NB likes to thank the Max Planck Institute for Solar System Research (MPS) for supporting his
visits at MPS during which a significant part of the work for this manuscript was performed.


\bibliography{pape,references}

\end{document}